\documentclass{acm_proc_article-sp}
\makeatletter
\newcommand{\removelatexerror}{\let\@latex@error\@gobble}
\newif\if@restonecol
\makeatother

\usepackage[ruled,vlined]{algorithm2e}
\usepackage{algorithmic} 
\usepackage{url}
\usepackage{subfig}
\usepackage{amsmath}
\usepackage{booktabs}
\usepackage[font=small,skip=0.5pt]{caption}
\usepackage{enumitem}
\makeatletter
\newcommand\subparagraph{%
  \@startsection{subparagraph}{5}
  {\parindent}
  {3.25ex \@plus 1ex \@minus .2ex}
  {-1em}
  {\normalfont\normalsize\bfseries}}
\makeatother
\usepackage{titlesec}
\let\subparagraph\relax % You don't want to use \subparagraph

\begin{document}

\title{MovePattern}
\subtitle{Interactive Framework to Provide Scalable Visualization of Movement Patterns}

\numberofauthors{3} %  in this sample file, there are a *total*
% 1st. author
\author{Kiumars Soltani$^{1,2}$,
Anand Padmanabhan$^{2,3,4}$, 
Shaowen Wang$^{1,2,3,4}$\\\vspace{6pt}
$^{1}${\em{Illinois Informatics Institute}}; \\\vspace{6pt}
$^{2}${\em{CyberGIS Center for Advanced Digital and Spatial Studies}}; \\\vspace{6pt}
$^{3}${\em{National Center for Supercomputing Applications (NCSA)}}; \\\vspace{6pt}
$^{4}${\em{Department of Geography and Geographic Information Science}}; \\\vspace{6pt}
\em{University of Illinois at Urbana-Champaign, Champaign, IL, USA}\\\vspace{6pt}
}

\permission{Permission to make digital or hard copies of all or part of this work for personal or classroom use is granted without fee provided that copies are not made or distributed for profit or commercial advantage and that copies bear this notice and the full citation on the first page. Copyrights for components of this work owned by others than ACM must be honored. Abstracting with credit is permitted. To copy otherwise, or republish, to post on servers or to redistribute to lists, requires prior specific permission and/or a fee. Request permissions from Permissions@acm.org.}
\conferenceinfo{IWCTS'15}{November 03-06, 2015, Bellevue, WA, USA}
\copyrightetc{ACM \the\acmcopyr}
\crdata{978-1-4503-3979-7/15/11...\$15.00.\\
DOI: http://dx.doi.org/10.1145/2834882.2834883}

% There's nothing stopping you putting the seventh, eighth, etc.
% author on the opening page (as the 'third row') but we ask,
% for aesthetic reasons that you place these 'additional authors'
% in the \additional authors block, viz.
% is the number that will appear on the first page PLUS the
% number that will appear in the \additionalauthors section.

\maketitle
\begin{abstract}
The rapid growth of movement data sources such as GPS traces, traffic networks and social media have provided analysts with the opportunity to explore collective patterns of geographical movements in a nearly real-time fashion. A fast and interactive visualization framework can help analysts to understand these massive and dynamically changing datasets. However, previous studies on movement visualization either ignore the unique properties of geographical movement or are unable to handle today's massive data. In this paper, we develop MovePattern, a novel framework to 1) efficiently construct a concise multi-level view of movements using a scalable and spatially-aware MapReduce-based approach and 2) present a fast and highly interactive web-based environment which engages vector-based visualization to include on-the-fly customization and the ability to enhance analytical functions by storing metadata for both places and movements. We evaluate the framework using the movements of Twitter users captured from geo-tagged tweets. The experiments confirmed that our framework is able to aggregate close to 180 million movements in a few minutes. In addition, we run series of stress tests on the front-end of the framework to ensure that simultaneous user queries do not lead to long latency in the user response.
\end{abstract}

\titlelabel{\thetitle.\quad}
\titlespacing*{\section}{0pt}{0.1\baselineskip}{0.1\baselineskip}
\titlespacing*{\subsection}{0pt}{0.1\baselineskip}{0.1\baselineskip}
\section{Introduction}
Movement datasets collected using location-aware devices (e.g. GPS tracking units and smart phones) have become an interesting type of Big Data attracting significant attention from many research communities (e.g., ecology, GIScience, and transportation) \cite{Bast:2014:RMV:2666310.2666404,CPE:CPE3287}. While movement data present tremendous opportunities for examining fine-grained geographical movements across multiple spatiotemporal resolutions, to effectively and efficiently derive knowledge from such data remains challenging \cite{journals/tvcg/ZinsmaierBDS12}.The challenges often lie in providing interactive and multi-resolution visualization framework that can handle massive datasets and large number of users \cite{2013-immens}. \par
\parskip 0pt
This paper provides a scalable system framework, called MovePattern, to demonstrate how interactive and multi-resolution visualization is achieved to enable a large number of users to study massive movement datasets. MovePattern consists of two key modules for data processing and visualization. The data processing module is centered on multiple scalable geospatial computing methods using Apache Hadoop \cite{Shvachko:2010:HDF:1913798.1914427}, which is capable of processing hundreds of millions of movements within a few minutes. Hadoop is an open-source software environment that supports distributed processing and storage of massive datasets, and has been used for solving a number of spatial analysis problems \cite{Liu:2010:MAG:1869692.1869695}. The data visualization module enables user interaction with the processed data to provide multi-resolution visualization of movements. We employ a vector-based visualization framework as opposed to pixel-based approaches which were used in previous work \cite{journals/tvcg/ZinsmaierBDS12}. In pixel-based approach the movement cannot be linked back to the original data, making it impossible for the user to get specific information about the nodes/edges after the visualization is produced. Therefore we use a vector-based approach to increase the analytical capabilities of the framework. Our multi-resolution approach to aggregate and summarize movements, address two main requirements for large-scale visualization system discussed in previous work \cite{2013-immens}. We provide ``perceptual scalability'' by providing an aggregated view of the movements in each spatial resolution thereby avoiding overwhelming user with too much information. In addition, we satisfy ``interactive scalability'' by providing a fast querying and visualization scheme. \par
\parskip 0pt
The MovePattern framework is evaluated using geo-tagged Twitter data \cite{CPE:CPE3287} and multiple computational experiments to demonstrate 1) significant scalability of MovePattern in aggregating and summarizing hundreds of millions of Twitter user movements; and 2) fast response to geographically distributed simultaneous queries from thousands of users. Results of these experiments confirm the ability of MovePattern for enabling a large number of users to perform interactive and multi-resolution visualization of movement data.
\setlength{\textfloatsep}{0.5pt}
\section{MovePattern Framework}
\subsection{Data Model}
In order to define the aggregation scheme, we first have to define a data model that demonstrates how the spatial domain is decomposed in our problem. To provide a easy-to-understand multi-level view of the  data, we adopted the hierarchical data cube approach discussed in previous studies \cite{CPE:Cao,Bast:2014:RMV:2666310.2666404} which divides the region of study into hierarchical uniform spatial grids. The cell size in each grid represents how detailed the information on the grid is and increases as we move to coarser resolutions. \par
\parskip 0pt
The multi-resolution modeling of data allows user to observe an overview of the data, while being able to drill down to get more focused information in a particular region. Therefore, the user is not going to be overwhelmed with too many movements being presented at once. Applying the data model into individual movement dataset will result in a multi-resolution spatial graph, where nodes are aggregated within the cells defined by the model and edges are aggregated movements among these nodes. The representative of the nodes are set to the centroid of all the points that lie into them.
\parskip 0pt
\subsection{Architecture}
\begin{figure}[ht!]
\centering
\includegraphics[scale=0.3]{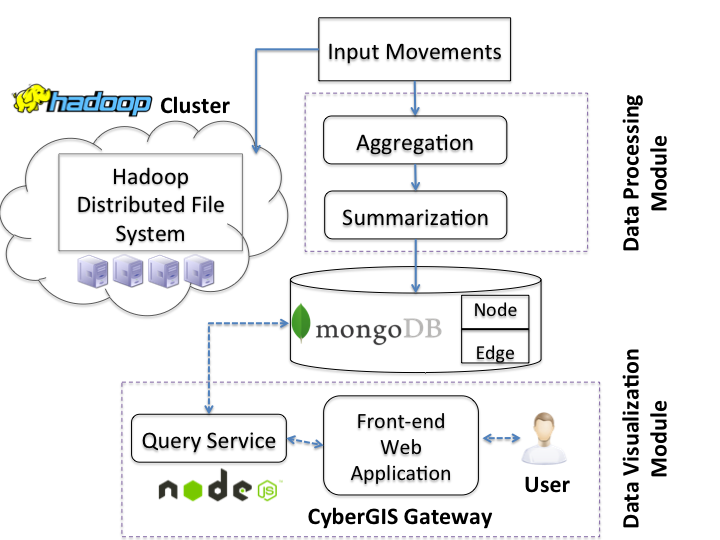}
\caption{The architecture of MovePattern}
\label{fig:arch}
\end{figure}

MovePattern needs to compute on accumulative movement data on regular basis to provide updated views for end users. Due to the massive data size, runtime user query processing on such large and increasing data is not practical. We thus separate the MovePattern data processing and interactive query and visualization into two modules:
\parskip 0pt
\begin{enumerate}[topsep=0pt,itemsep=0.05em]
\item The \textbf{\textit{data processing module}} is responsible for processing raw movement data, aggregating them to form the multi-level view and finally summarizing it.
\item The \textbf{\textit{data visualization module}} interacts with the user to transform their query into the visualization result. This component is called on-demand based on incoming user requests.
\end{enumerate}
Figure \ref{fig:arch} illustrates the overall architecture of MovePattern. \par
\parskip 0pt
\parskip 0pt
Since the data processing module requires handling a large number of movements, it is crucial to design and implement this module using a scalable framework. The MapReduce based design allows us to implement this module using Apache Hadoop, an open-source implementation of MapReduce \cite{Dean:2004:MSD:1251254.1251264}. MovePattern benefits from Hadoop by distributing the massive input data into multiple nodes, scheduling parallel tasks among multiple machines and enabling seamless scaling by growing the cluster. In addition, Hadoop automatically re-route the computation in case one of the machines fail or perform slowly. In section 3, we discuss how we can divide the computation evenly among multiple machines using Hadoop capabilities.
\section{Data Processing Module}
The data processing module transforms the raw movements data into a concise multi-resolution view of collective movements. The module includes three main steps: 1) aggregate the points and movements to form a multi-resolution spatial graph; 2) find ``significant'' nodes in each aggregation resolution; 3) remove the movements that are not associated with significant nodes. A spatial partitioning scheme is applied to the input data for all three steps, which is crucial to improve the performance of all three steps.
\parskip 0pt
\subsection{Spatial Partitioning}
One of the most significant issues in MapReduce applications is to deal with skew \cite{Kwon:2012:SMS:2213836.2213840}. While in Hadoop, the input data is evenly distributed among the mappers, based on the configuration variable $dfs.block.size$, the applications can still suffer considerably from skew among reducers. The reason for existence of such skew is the inability of Hadoop to dynamically balance the load among reducers. Particularly for spatial application, many of the real-world data are highly skewed and therefore a custom partitioner is required to balance the computation in Hadoop. To address this issue, we pre-process a small sample of the movement dataset to form a spatial indexing scheme (there is no specific order assumed for the input data). \par
\parskip 0pt
To build the spatial partitioning scheme, we use the recursive bisection approach that partitions a space into a set of rectangles \cite{1676942}. In this method at each step, we divide the region into two sets in the way that it minimizes the imbalance. One main variation of our approach from the original recursive bisection method is that instead of alternating the cut axis at each level, we choose the axis that gives us a better balance among two sets. After building this partitioning, we load the partitions at the start of each mapper, using the initial setup function (each mapper processes multiple data blocks, therefore for a series of data block, we only load the partitions once). Then for each movement the mapper determines the partition based on the loaded partitioning scheme and the movement's source latitude and longitude (the aggregation is done using the source node). To increase the performance of the lookup process, we index the partitions using R-tree \cite{Guttman:1984:RDI:602259.602266} to leverage its capability to provide fast ``\textit{contains}'' operation. 
\parskip 0pt
\subsection{Movement Aggregation}
MovePattern aggregates the input data in multiple spatial resolutions to provide better insights on collective analysis of individual's movements. As mentioned in Section 2.1, MovePattern adopts the hierarchical data cube approach discussed in previous work \cite{CPE:Cao,Bast:2014:RMV:2666310.2666404} which divides the region of study into hierarchical uniform spatial grids. While this approach provides a general spatiotemporal cube with efficient query time, the cube generation is still time-consuming \cite{CPE:CPE3287}. Therefore, we designed a MapReduce algorithm to efficiently generate the cube. \par
\parskip 0pt
The outline of MovePattern spatial aggregation algorithm is explained in Algorithms \ref{alg:aggrmapper} (mapper) and \ref{alg:aggrreducer} (reducer), where $L$ is the number of resolution levels. The uniform structure of our grid enables implementing $getCellId$ in $O(1)$ without performing any pre-processing on the dataset. One final improvement to reduce the graph size is filtering edges based on maximum edge distance allowed in each spatial resolution. To better explain the distance-filtering scheme, suppose the user zooms into an area around New York City to explore movement patterns. In this case, movements from New York City to Los Angles are not useful to be visualized since they are out of the area of interest. We use this principle and define a maximum threshold for distance of nodes for each resolution. \par
\parskip 0pt
To further speed up the processing, we use in-mapper combiner as opposed to Hadoop built-in combiner. Using this technique we can make sure that 1) the combiner is being called for all the processed keys in each mapper and 2) avoid possible intermediate spill to disk before the map process finishes. 
\parskip 0pt 
\begin{algorithm}[h]
\SetKwProg{Fn}{}{}{}\SetKwFunction{mapper}{function mapper}
\Fn{\mapper{movement}}{
$s \gets movement.source$ \;
$t \gets movement.target$ \;
\For{$l = 1...L$}{
$id1 \gets getCellId(s, l)$ \;
$id2 \gets getCellId(t, l)$ \;
$merge(graphs[l], id1, Node(s, 1))$ \;
$merge(graphs[l], id2, Node(t, 1))$ \;
\If{$d(s,t) < threshold(l)$}{
$merge(graphs[l], id1,  Edge(id1, id2, 1))$ \;
}
}
}
\SetKwProg{Fn}{}{}{}\SetKwFunction{cleanup}{function cleanup}
\Fn{\cleanup{}}{
\For{$l = 1...L$}{
\ForEach{$key$ in graphs[l]}{
$p \gets rtree.search(key)$ \;
$emit(p, [l, graphs[l].get(key)])$ \;
}
}
}
\caption{Mapper for spatial aggregation}
\label{alg:aggrmapper}
\end{algorithm}
\parskip 0pt
\begin{algorithm}[h]
\SetKwProg{Fn}{}{}{}\SetKwFunction{reducer}{function reducer}
\Fn{\reducer{key, Node[] values}}{
\ForEach{$value$ in values}{
$merge(graphs[value.level], value.id, value)$ \;
}
\For{$l = 1...L$}{
\ForEach{$key$ in graphs[l]}{
$x \gets graphs[l].get(key)$ \;
$write("Nodes", x.level, x.coor, x.count)$ \;
$write("Edges", x.neighbors)$ \; 
}
}
}
\caption{Reducer for spatial aggregation}
\label{alg:aggrreducer}
\end{algorithm}
\parskip 0pt
The result of aggregation step is a hierarchical spatial graph, which includes aggregated analytical measures which are computed in the $merge$ function (e.g. count, number of users, average travel time, etc.).
\parskip 0pt
\subsection{Node Summarization}
While multi-resolution spatial aggregation provides a generalized view of data, it can be still too large to convey any clear patterns to users in a visualization interface. For instance, if we divide North America into $512km \times 512 km$ cells, we will end up with 468 grid cells which can have up to 109278 edges among them. Therefore, even for very coarse-level view of data, we get a very cluttered graph; hence the result from node aggregation step needs to be summarized for a less cluttered visualization. \par
\parskip 0pt
Our summarization technique filters less ``significant'' nodes (grid cells) by assigning them a score, reflecting how large their degree is comparing to neighboring cells. Then by comparing the score to a pre-defined threshold we can decide whether to keep or drop a node from the graph. As previous research has pointed out \cite{CPE:CPE3287} geographical distribution of real-life location-based data is highly skewed, with a small number of places contributing most of the activities. Therefore, the definition of ``significant'' nodes should be localized to their surrounding sub-regions as opposed to using the same significance measure for the whole region of study. \par
\parskip 0pt
The local neighborhood of point $p$ is defined as $\{q \in P : d(p,q) < r\}$. Here $P$ is the set of all points in the graph (in the same spatial resolution as $p$), $d(p,q)$ is the great-circle distance between $p$ and $q$ and $r$ is the neighborhood radius. By reducing $r$ we will have a more strict definition of a neighborhood which will lead to having more points in the final graph. The value of $r$ can be adjusted for each spatial resolution. \par
\parskip 0pt
To model this problem using MapReduce, we have to be cautious to avoid unnecessary communication among different nodes. In the naive approach each node send their information to every other node, and help them find about their neighborhoods. However, this will lead to very expensive communication overhead. Instead we take advantage of the partitioning scheme that was built in the initial stage of the data processing module to prune many choices and end up with only a small set of cells as neighbor candidates. The uniform structure of the grid enables us to easily extract the set of cells, which are in a certain distance from the current cell. The outline of this MapReduce based approach is explained in Algorithms \ref{alg:summapper} and \ref{alg:sumreducer}. The input of this job is the "Node" output of the aggregation step.
\parskip 0pt
\begin{algorithm}[h]
\SetKwProg{Fn}{}{}{}\SetKwFunction{getnp}{function getNeighborPartitions}
\Fn{\getnp{node, level}}{
$offset \gets cell\_len[level] \times r[level]$ \;
\textit{result-set} $\gets rtree.findNeighbors($ \par
\hskip\algorithmicindent $node.coor, offset)$ \;
\Return{result-set} \; 
}
\SetKwProg{Fn}{}{}{}\SetKwFunction{mapper}{function mapper}
\Fn{\mapper{node}}{
\ForEach{$p$ in $getNeighborPartitions(node, node.level)$}{
$emit(p, node)$ \;
}
}
\caption{Node Summarization Mapper}
\label{alg:summapper}
\end{algorithm}
\parskip 0pt
\begin{algorithm}[h]
\SetKwProg{Fn}{}{}{}\SetKwFunction{reducer}{function reducer}
\Fn{\reducer{key, Node[] values}}{
\For{$l = 1...L$}{
$rtrees[l] \gets $ new rtree() \; 
}

\ForEach{$node$ in values}{
\If{!rtrees[node.level].contains(node)}{
$rtrees[node.level].add(node)$ \;
}
}
\For{$l = 1...L$}{
\ForEach{$node$ in rtrees[l]}{
$offset \gets cell\_len[node.level] \times r[node.level]$ \;
\textit{result-set} $\gets rtrees[node.level].findNeighbors($ \par
\hskip\algorithmicindent $node.coor, offset)$ \;
$rank \gets $ percentile rank of $node.count$ in \textit{result-set} \;
\If{rank > threshold}{
$write("SummaryNode", $ \par
\hskip\algorithmicindent $node.id, node.level, node.coor, node.count)$ \;
}
}
}
}
\caption{Node Summarization Reducer}
\label{alg:sumreducer}
\end{algorithm}
\parskip 0pt
\subsection{Edge Filtering}
The node summarization step provides us with a list of significant nodes. The next step is to filter the edges among significant nodes to build the aggregated summarized final graph. The trivial solution for filtering edges is to perform a join on the aggregated edges and the list of summarized nodes. However the join operation on such large data can be quite time-consuming. Therefore we propose a fast probabilistic method that takes advantage of Bloom Filters \cite{Bloom:1970:STH:362686.362692} to filter the list of edges. Bloom filter is a well-known data structure that stores a series of entries in an space-efficient fashion and can be used to test whether an entry is a member of the data structure or not. Bloom filter uses $k$ independent hash functions and a binary array of length $m$ to predict membership of an element with probability $p$. The parameters can be tuned considering space limitation and desired false positive limit. The key point about the probabilistic nature of Bloom filter is that while false positive matches may occur, there will be no false negatives. Therefore, we can guarantee that no significant edge will be removed. \par
\parskip 0pt
After building the bloom filter for the summarized nodes (one filter for each level), we run a simple MapReduce job to go through the list of edges and check whether both source and target of edges can pass the membership test of Bloom filter. If the edge passes the test, then we will write them to the final list of edges. The bloom filters are shared among mappers using the distributed cache capability in Apache Hadoop.

\section{Data Visualization Module}
The data visualization module of MovePattern is responsible for managing interactions with users and visualizing movements by consuming the output produced by the data processing module. The module consists of a front-end end web application and a query service. \par
\parskip 0pt
The query service contacts the MongoDB database\footnote{https://www.mongodb.org/} to get processed movements based on user request. We store the result of processing module as an spatiotemporal data cube in MongoDB where node collections are geographically indexed to perform fast bounding box queries. The query service is implemented using NodeJS\footnote{https://nodejs.org/} in an asynchronous fashion and therefore the user requests do not block each other. This is a crucial factor in designing interactive frameworks where the status of one user request should not affect other users requests. \par
\begin{figure}[ht!]
\centering
\includegraphics[scale=0.2]{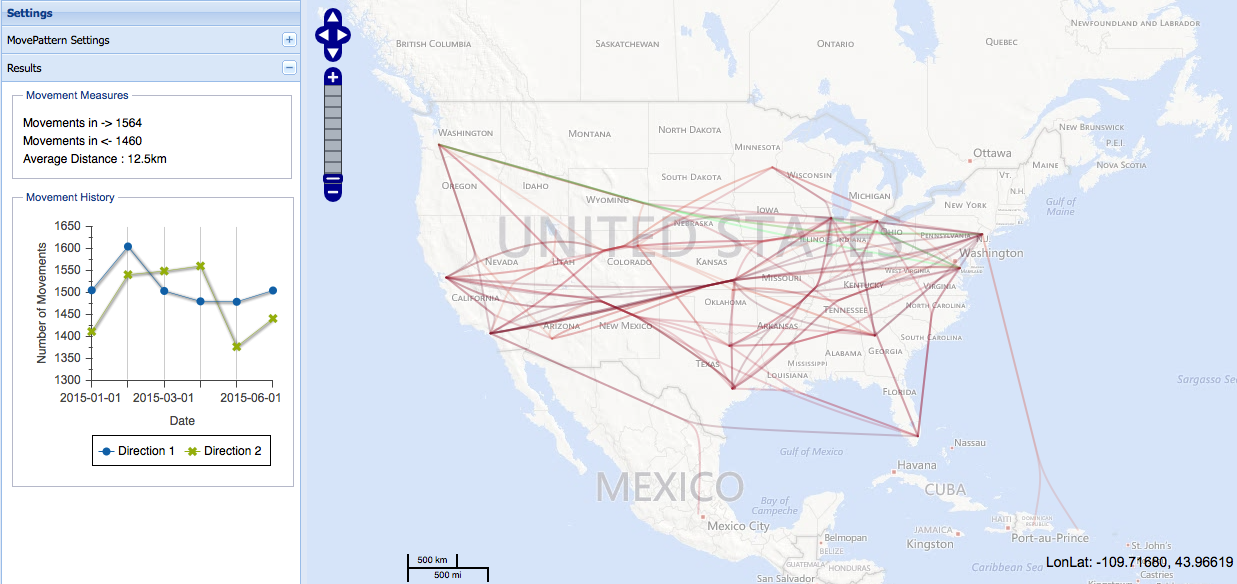}
\caption{MovePattern front-end web application}
\label{fig:app}
\end{figure}

\parskip 0pt
The front-end web application serves as a gateway to the capabilities of the MovePattern framework. The web application enables user to select a region by panning and zooming in/out and then contacts the query service to obtain the subgraph enclosed in the selected region. Moreover the users can specify the time period and level of details to customize the visualization result. This application has been integrated into CyberGIS Gateway\cite{6702694} - an online environment for a large number of users to perform computing and data-intensive, and collaborative geospatial problem solving. An overlay of the application is illustrated in Figure \ref{fig:app}.\par
MovePattern employs a vector-based visualization as opposed to a pixel-based visualization. By using vector-based visualization, we are able to 1) store attribute information for each node/edge and 2) perform fast client-based customization (without  additional client-server interaction). \par
\parskip 0pt
To further reduce the clutter of the visualization we visualize the movements using the MINGLE edge bundling algorithm \cite{5742389}. MINGLE is a scalable edge bundling approach that focuses on minimizing the virtual ink required to draw a graph. This algorithm avoids all-to-all similarity comparison of edges by forming an edge proximity graph (modeling edges as points in 4-dimensional space). We use a javascript implementation of MINGLE\footnote{https://github.com/philogb/mingle} to perform a client-side visualization, hence reducing the load on our query service when multiple simultaneous requests are made.
\section{Experiments} 
In this section we present a set of experiments, which were conducted to evaluate the scalability of the MovePattern framework with increase of both graph size and number of concurrent user queries. The experiments have been conducted on the ROGER\footnote{Resourcing Open Geospatial Education and Research} supercomputer in the National Center for Supercomputing Applications. For the Hadoop cluster, we used 8 nodes each includes 2 Intel Xeon E5-2660 2.6 GHz CPU (20 cores total), 256 GB of memory and 800GB of SSD hard drive. Both web server and database instances are also launched as OpenStack virtual machines on ROGER. For the NodeJS webserver, we use a virtual machine with 4 cores of Intel Xeon E5-2660 2.6 GHz CPU and 8GB of memory. The MongoDB database is designed as 4-node replica set with 2 cores and 4GB of RAM. For all the experiments we performed 3 separate runs and averaged the result for the final measure.
\subsection{Twitter-based Movements} 
We captured user movements based on their geo-tagged tweets for the period of three months, starting August 1st 2014 through October 31st 2014. The tweets were collected and geo-referenced based on the Twitter Streaming API \cite{CPE:CPE3287} and movements generated by forming a spatiotemporal trajectory for each user. To divide the data into multiple spatial resolutions, a hierarchical uniform grid is formed with 10 levels, representing different level of details for the North America continent. At the finest level, a uniform grid with cell size of 30 arc seconds (approximately 1 $km$ $\times$ 1 $km$) is formed and the cells are iteratively merged to form the uniform grids on the higher level. The merging operation is done in an exponentially increasing fashion forming cell sizes of 2, 4, \dots, 512 $km$. In addition, each cell is presented using the centroid of all the points (location of tweets) within the cell. \par
\parskip 0pt
Table \ref{tbl:inputdata} shows the number of tweets, unique users and movements in the collected dataset.
\begin{table}[ht!]
\centering
\begin{tabular}{lcccc|}
Duration & Tweets & Users & Movements & Size(GB) \\
\hline
1-month & 107M & 2.2M & 76M & 2.77 \\
2-month &  205M & 2.9M & 136M & 5.43 \\
3-month &  368M & 3.5M & 179M & 9.88 \\
\hline
\end{tabular}
\caption{Input data statistics} 
\label{tbl:inputdata}
\end{table}
\parskip 0pt
\subsection{Data Processing Module Performance} 
We first demonstrate the advantage of our partitioning scheme by comparing the load on each reducer. Then we present the performance of aggregating and summarizing three datasets using the elapsed time and average mapper/reducer time. \par
\parskip 0pt
Table \ref{tbl:balance} shows the statistics on reducers load when aggregating data for the 3-month dataset. This result confirms that using the partitioning scheme, described in section 3.1, we can divide our study space into multiple regions with similar computational load. \par
\parskip 0pt
\begin{table}[!ht]
\centering
\begin{tabular}{lccccc|}
\# of Reducers & Avg & Std & Min & Max \\
\hline
8&5,853,967&120,348&5,709,732&6,037,848 \\
16&2,926,984& 73,670&2,821,192&3,096,001 \\
32&1,463,492&58,673&1,344,301&1,609,537 \\
64&731,746&45,466& 654,939&846,538 \\
\hline
\end{tabular}
\caption{The reducers' load statistics in number of mapper outputs} 
\label{tbl:balance}
\end{table}
\parskip 0pt
The next experiments focus on the performance of spatial aggregation and summarization methods on the three test datasets. For these experiments we set the HDFS block size to 64MB (this factor determines the launched map tasks for each dataset) and the node summarization threshold to 80\%. Table \ref{tbl:result} shows the result of running spatial aggregation and summarization methods on the three datasets with 4, 8, 16, 32, 64 and 128 reducers. Determining number of reducers for a dataset is challenging since we face a trade-off between having more concurrent tasks and the additional overhead that arises from having too many reducers. Therefore we run the experiments for different numbers of reducers to determine which one give us the best performance. \par
\parskip 0pt
For the largest input, which consists of over 178 million movements the overall processing time is 94 seconds. As we can see in Table \ref{tbl:result}, by using different number of mappers and reducers for each dataset, the computing time only marginally increased as we move to larger datasets. \par
\parskip 0pt
\begin{table}[ht!]
\setlength{\tabcolsep}{1.5pt}
\begin{tabular}{lrrrrrr}\toprule
Dataset &\multicolumn{2}{c}{\textbf{Aggregation}}&\multicolumn{2}{c} {\textbf{Summarization}}&\multicolumn{2}{c}{\textbf{Filtering}}
\\\cmidrule(r){2-3}\cmidrule(r){4-5}\cmidrule(r){6-7} 
& reducers & time & reducers & time & reducers & time \\\midrule
1 month & 32 & 39 & 16 & 25 & 4 & 18 \\
2 months & 32 & 43 & 32 & 26 & 4 & 19 \\
3 months & 64 & 47 & 32 & 28 & 8 & 19
\\\bottomrule
\end{tabular}
\caption{The result of data processing module (in secs)}
\label{tbl:result}
\end{table}
\parskip 0pt
\begin{figure}[ht!]
\centering
\includegraphics[scale=0.5]{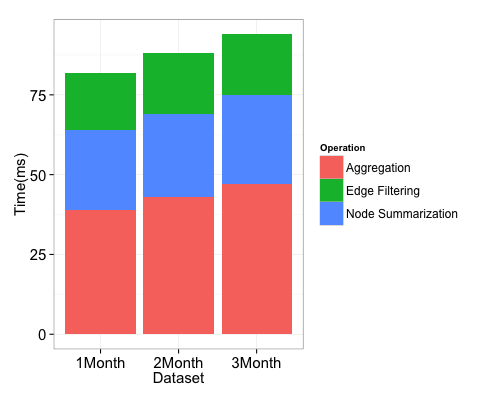}
\caption{The overall time of data processing module}
\label{fig:overall}
\end{figure}
\parskip 0pt
Finally, we present the effect of aggregation and summarization on the size of the graph. Table \ref{tbl:effect} shows how the aggregation and summarization reduce the number of nodes/edges in the graph. This abstraction reduces the time necessary for processing and visualizing the movements as well as reduces the clutter of the visualization by focusing on the most significant movements in each level. 
\parskip 0pt
\begin{table}[h]
\begin{tabular}{lccc}\toprule
Dataset & Nodes & Edges \\
1 month & $2.67M \rightarrow 939K$ & $33.98M \rightarrow 4.66M$ \\
2 months & $3.18M \rightarrow 1.08M$ & $49.27M \rightarrow 6.88M$ \\
3 months & $3.48M \rightarrow 1.16M$ & $59.55M \rightarrow 8.44M$ \\
\bottomrule
\end{tabular}
\caption{Effect of aggregation and summarization on graph size}
\label{tbl:effect} 
\end{table}
\parskip 0pt
\subsection{Interactive Scalability}
To evaluate interactive scalability, we simulate two categories of queries that represent two most common query patterns of users:
\parskip 0pt
\begin{enumerate}[topsep=0.1pt, itemsep=0pt]
\item \textbf{Population Query Pattern}: Queries are distributed in a weighted fashion around the region of study (here North America) with more queries on sub-regions with higher number of tweets.
\item \textbf{Hotspot Query Pattern:} Queries are focused on a specific relatively small region resembling occasional situations which an outburst will lead to large number of focused access. For instance, a political visit or a natural disaster can attract user attentions to a certain area.
\end{enumerate} \par
\parskip 0pt
The underlying assumption in both access patterns is that the framework is likely to get more queries from regions where there are more tweets. Therefore, we generated the query bounding box according to a sample of tweets, where more crowded regions are more likely to be presented in the sample. In addition, the spatial resolution is randomly chosen in a uniform fashion from  $\{1,2,...,10\}$. \par
\parskip 0pt
For our experiments, we generated 3 sets of 2000 queries for overall and focused query pattern. Then Apache JMeter \footnote{\url{http://jmeter.apache.org/}}, a load testing tool, is used to send this queries to MovePattern with different rates of queries per second and the response time is measured. We launched 2000 queries in the duration of 50, 75 and 100 seconds, on the 3-month dataset. After performing aggregation and summarization, this dataset contains over 1.16M nodes and 8.44M edges. Table \ref{tbl:interactive} shows statistics (average, median and 90\% percentile) on the response time of queries for both normal and focused patterns. The result shows that MovePattern can sustain relatively large number of simultaneous queries, each based on different resolutions and regions. \par
\parskip 0pt
\begin{table}
\subfloat{\begin{tabular}{lcccc}
\multicolumn{4}{r}{\textbf{Population Pattern}}
\\\cmidrule(r){2-5}
& Duration(s) & Average & Median & 90\% Percentile \\\midrule
& 50 & 42&29&96 \\
& 75 & 34&23&79 \\
& 100 & 31&21&72
\\\bottomrule
\end{tabular}}
\hfill
\subfloat{
\begin{tabular}{ccccc}
\multicolumn{4}{r}{\textbf{Hotspot Pattern}}
\\\cmidrule(r){2-5}
& Duration(s) & Average & Median & 90\% Percentile \\\midrule
& 50 & 36&24&82 \\
& 75 & 31&21&69 \\
& 100 & 28&19&63
\\\bottomrule
\end{tabular}
}
\caption{Stress test for 2000 queries on 3-month dataset (in ms)}
\label{tbl:interactive}
\end{table}
\section{Concluding Discussions}
In this paper we introduced MovePattern, a scalable framework for interactive and multi-resolution visualization of massive movement data. MovePattern uses a suite of MapReduce algorithms, implemented in Apache Hadoop, to process hundreds of millions of movements in matter of minutes. These algorithms aggregate the movements at multiple spatial resolutions and then summarize them to only keep the most significant ones. The processed movements will then be accessible through a highly interactive web application that employs a vector-based visualization technique to link the movements with their underlying characteristics. We evaluated the framework using the Twitter user movements using three months of geo-tagged tweets. MovePattern were able to aggregate and summarize more than 178 million movements in 94 seconds, while keeping the query latency for user interaction to under 100ms.
\section*{ACKNOWLEDGMENTS}
This material is based upon work supported in part by the National Science Foundation (NSF) under grant numbers: 1047916, 1354329 and 1443080. The work used the ROGER supercomputer, which is supported by NSF under grant number: 1429699. The authors would also like to thank the members of the CyberInfrastructure and Geospatial Information Laboratory (CIGI, http://cigi.illinois.edu/) for their insightful comments and discussions

\bibliographystyle{abbrv}
\bibliography{sigproc-sp}  % sigproc.bib is the name of the Bibliography in this case

\balancecolumns
% That's all folks!
\end{document}